# ESA Voyage 2050 White Paper

## OWL-MOON:
## Very high resolution
## spectro-polarimetric interferometry and imaging
## from the Moon:
## exoplanets to cosmology

**Contact :**

Jean Schneider

Observatoire de Paris

France

jean.schneider@obspm.fr

Joseph Silk

IAP (France)/John Hopkins University (USA)

joseph.silk@physics.ox.ac.uk

Farrokh Vakili

Observatoire de la Côte d'Azur, Nice

France

farrokh.vakili@oca.eu

**Summary**

We address three major questions in astronomy, namely the detection of biosignatures on habitable exoplanets, the geophysics of exoplanets and cosmology. To achieve this goal, two requirements are needed : 1/ a very large aperture to detect spectro-polarimetric and spatial features of faint objects such as exoplanets, 2/ continuous monitoring to characterize the temporal behavior of exoplanets such as rotation period, meteorology and seasons. An Earth-based telescope is not suited for continuous monitoring and the atmosphere limits the ultimate angular resolution and spectro-polarimetrical domain. Moreover, a space telescope in orbit is limited in aperture, to perhaps 15 m over the next several decades. This is why we propose an OWL-class lunar telescope with a 50-100 m aperture for visible and infrared (IR) astronomy, based on ESO's Overwhelmingly Large Telescope concept, unachievable on Earth for technical issues such as wind stress that are not relevant for a lunar platform. It will be installed near the south pole of the Moon to allow continuous target monitoring. The low gravity of the Moon will facilitate its building and manoeuvring, compared to Earth-based telescopes. As a guaranteed by-product, such a large lunar telescope will allow Intensity Interferometric measurements when coupled with large Earth-based telescopes, leading to pico-second angular resolution.

# 1 Introduction

The search for life on exoplanets is humanity's greatest challenge. One cannot imagine a more inspiring goal. The current focus for the next decade will be the search for super-Earths, revealed by Kepler to predominate among exoplanets in our interstellar neighborhood. Larger than Earth and smaller than Neptune, some may support liquid water oceans and Earth-like biology. This will take larger and more ambitious telescopes.

Currently, our primary hope is the LUVOIR mission, a 8-15m aperture telescope to be launched to L2, with UV-NIR imaging and spectoscopic capability over 200-2000 nm. LUVOIR, to serve the astronomical community in the 2030s, will characterise habitable zone exoplanets and search for biosignatures, as well as explore the formation of the first galaxies and quasars.

The LUVOIR mission has identified a total of 50 habitable zone rocky exoplanets within 25 pc of the sun that are suitable for biosignature follow-up. We demonstrate that a large lunar telescope can triple the survey volume and perhaps provide the first serious opportunity we will have within the next few decades for assessing the complexity of extrarrestrial life on earth-like exoplanets. It is a unique opportunity, that we should exploit now that the international space agencies are seriously considering a return to the Moon.

The next frontier in human exploration of space is to extend our current capabilities to rocky planets in the habitable zones of their host stars. We will not only identify but follow-up with detailed spectrocsopic polarimetry and imaging those "pale blue dots" within tens of pc of the sun. There are thousands of candidates accesible to a large lunar telescope. We can determine not merely whether habitable, Earth-like conditions are common or rare on nearby worlds, but and probe our neighbouring exoplanets for signs of life.

It is hugely challenging. The Earth is 10 billion times fainter than the Sun and orbits close to its host star : viewed from 100 pc, the separation is only 0.01 arc-sec. But this is well above the diffraction limit of a very large lunar telescope. We can study exoplanet atmospheres from a lunar platform, where there is no atmosphere to confuse our signal. Telescope size simultaneously guarantees a large number of earth-like targets. We cannot fail, if the will is there to develop known technology, with the aid of robotic resources in deep icy craters near the south pole, in permanent darkness and where temperatures approach 30K, with adjacent crater rims in perpetual sunlight to provide solar power.

These environments are the prime focus of future lunar exploration, and are equally prime sites for lunar telescopes. We must develop a persuasive agenda for ESA for telescope construction in order to achieve humanity's greatest science goals, an agenda so persuasive that it cannot be ignored. The technology is challenging, but no more so than that required for ELT or LUVOIR. Construction and deployment provide challenges that will undoubtedly become feasible as the new age of lunar exploration progresses.

In addition to these scientific perspective, we develop a novel approach of Intensity Interferometry (II). While previous II projects were based on large photon collectors with bad surface mirror quality like the Cherenkov Telescope Array, we take the advantage of the very narrow PSF 1micron/50 m = 4 milli arcsec to split the signal into several thousand spectral channels thanks to integrated optics. We explain this in details in section 4.

*1a. Justification of a large aperture*

Let us first justify an aperture as large as 50m or more. One of our main science objectives is the characterization of exoplanets and biosignatures. There are about ten potentially habitable planet candidates up to 10 pc. But there is no guarantee that even a single one will present biosignatures. We must enlarge the sample and go up to say 40 pc. An Earth-sized planet at 1 AU from a G star has a planet/star brightness ratio of $3.10^{-9}$ for an albedo of 0.3. Thus, for a 8th magnitude star, it means a 32nd magnitude target. For 1 nm spectral resolution spectroscopy needed to detect atomic and molecular emission lines, consider the goal of 1000 photons detected in 3 hours. This needs a 50 m telescope. To detect 500 photons in the bottom of absorption lines having a depth 10 times the continuum in 3 hours, one would need a 100 m telescope. OWL was designed to be 100 m, but was abandoned for technological reasons to be replaced by the 39m ELT, now under construction on Cerro Armazones in the Atacama Desert of northern Chile.

There are no such limitations in the lower gravity and wind-free conditions on the lunar surface. At an ambient stable thermal environment near 30K, allowing one to reach some 7K with passive cooling, ideal for astronomy through the FIR. With no atmosphere, the observing frequency window provides space-like conditions, with the advantage of a very large aperture on a stable platform (Lester et al. 2004). In addition, contrary to standard interferometry, Intensity Interferometry is not sensitive to lunar seismology. Here we focus on conventional telescope designs (Moretto, 2016); other concepts should be further explored, including cryogenic liquid mirror designs (Angel et al. 2008), although the later is sensitive to lunar seismology.

## 2. Science Objectives

### 2a. Characterization of exoplanets

*2a(i). Geophysics of exoplanets*

After the mass of a planet, three major global characteristics are its radius, its albedo and its period of rotation. The latter is easily inferred from its continuous photometric monitoring (Ford et al. 2001). The radius is easily measured for transiting planets. But the majority of them are not transiting and the only observable by direct imaging is the total flux = albedo x area, where the albedo is a function of wavelength and time and the area is a priori unknown . Nevertheless, one can disentangle this degeneracy with generalized atmospheric circulation models if the monitoring of the flux is continuous (*e.g.* Read et al. 2018). Beyond these global characteristics, continuous spectro-photometric monitoring will reveal oceans and continents of planets and their relative area (Ford et al. 2001). Also, McCullough (2006) and Williams & Gaidos (2008) have proposed to detect the specular reflection of the parent star on oceans of exoplanets. The follow-up of the specular reflection will help to shape the non-reflective parts of the planets, that is their continents.

### 2a(ii). Atmospheres

Beyond the molecular composition of its atmosphere, spectroscopy also provides at short wavelengths, for clear skies, the amount of Rayleigh scattering and thus the atmospheric pressure.

### 2a(iii). Surfaces

For rocky planets with clear skies, the averaged mineral composition will follow from the comparison of the spectrum with spectra of minerals databases such as the CalTech Mineral Spectroscopy Server at http://minerals.gps.caltech.edu/ .

### 2a(iv). Polarization

The light of the parent star reflected by the planet is linearly polarized by the atmosphere molecules and aerosols and by the surface (if it is visible), whereas the star is generally unpolarized. Therefore polarimetry helps to disentangle the planet from unavoidable stellar speckles due to the imperfections of the optics. In addition, the degree of polarization varies along the planet orbit with the phase angle of the planet leading to a confirmation that the polarization is due to the planet and not to the instrumentation.

### 2a(v). Biosignatures

There are two types of spectral biosignatures : out of chemical equilibrium gases due to biological activity (such as photosynthetic oxygen on Earth) and direct detection of organisms (such as vegetation on Earth (O'Malley & Kaltenegger 2019)). There are continuing debates on the robustness of gaseous biosignatures (*e.g.* Chan et al. 2019). It is important to accumulate measurements in time and to couple them with the orbital phase and the stellar spectrum. To consolidate direct detection of « vegetation » it will be important to monitor it continuously in time in order to correlate with seasons. In order to be a reliable « vegetation » candidate, a spectral signature must be absent from databases of mineral spectra such as the CalTech Mineral Spectroscopy Server.

In addition to spectroscopy, circular polarimetry can search for direct detection of extrasolar « vegetation » analogs (Patty et al. 2019).

Although we do not propose this as a goal, we are also open to the exploration of possible technosignatures (e.g. Wright 2019). An example could be spectral and temporal features of optical characteristics due to technological activities.

*2a(vi). Mountains and volcanos on planets*

Whereas several of the above science topics can be addressed with a single OWL on the Moon, some questions require the extremely high angular resolution afforded by an Earth-Moon Intensity Interferometer. Once an OWL-type telescope is installed on the Moon, or even a 10 m lunar precursor, one could readily address optical Intensity Interferometry with unprecedented baselines and angular resolution. For instance it could measure the heights of mountains on transiting exoplanets. This is an important problem for the geophysics of planets. V. Weisskopf (1975) has shown that there is a relationship between the maximum height of mountains on a planet and its mass and the mechanical characteristics of its crust.

The issue of their detectability has already been addressed for transiting planets (McTier & Kipping 2018). Here we propose a significant improvement, Based on the principle of the detection of the silhouette of ringed planets by Intensity Interferometry as developed by Dravins (2016). With a 60m resolution at the 1.4 pc distance of alpha Cen, for transiting planets, mountains will appear at the border of the planet silhouette during the transit. These observations will require very long exposures. During the exposure, the planet is rotating around its axis, leading to a washing-out of the features we are looking at. But the planet rotation period will be well known from the periodicity of its photometric data (Ford et al. 2001). Therefore, the mountain silhouette will appear in a 2D Fourier transform of long series of short exposure images at the planet rotation frequency. Moreover, volcanos can be detected as a temporary excess of red emission of the planet.

*2a(vii). Image of stellar glint on exo-oceans*

It will be possible to make an image of specular reflection of the parent star on its oceans. Indeed, from McCullough (2006), one gets for the flux received from the glint of the ocean of an Earth-sized planet around a solar-type star at 10 pc and for an ocean albedo 6 %, 7 photons/sec with 30 m telescopes. The monitoring of this image would reveal the contours of the continents.

Several other future exoplanet science objectives can be found in Schneider (2018).

### 2a(viii). Exoplanet Counts

We examine the exoplanet science capabilities of OWL-MOON. Comparing this to a proposed 15 m space telescope (LUVOIR interim study 2018) with capabilities of observing exoplanet spectra of ~50 Earthlike planets, defined to have rocky cores and be in habitable zones of the host stars, around main sequence stars within 25 pc and over the course of a 25 year mission. Using the scaling with telescope diameter (Stark et al. 2014) for noise dominated by zodiacal light, N proportional to D to the power of 1.8, we find that 3000 planets could be imaged for a 50-100 m telescope. For photon count-limited noise, the number scales as N proportional to D to the power of 1.2, yielding instead 1,000 planets for the 50-100m aperture design. Such a yield is a huge statistical gain over LUVOIR and brings significantly enhanced chances of finding atmospheric biosignatures for exoplanets within a search radius of 250 pc. The advantages of aperture size are shown in Fig.1.

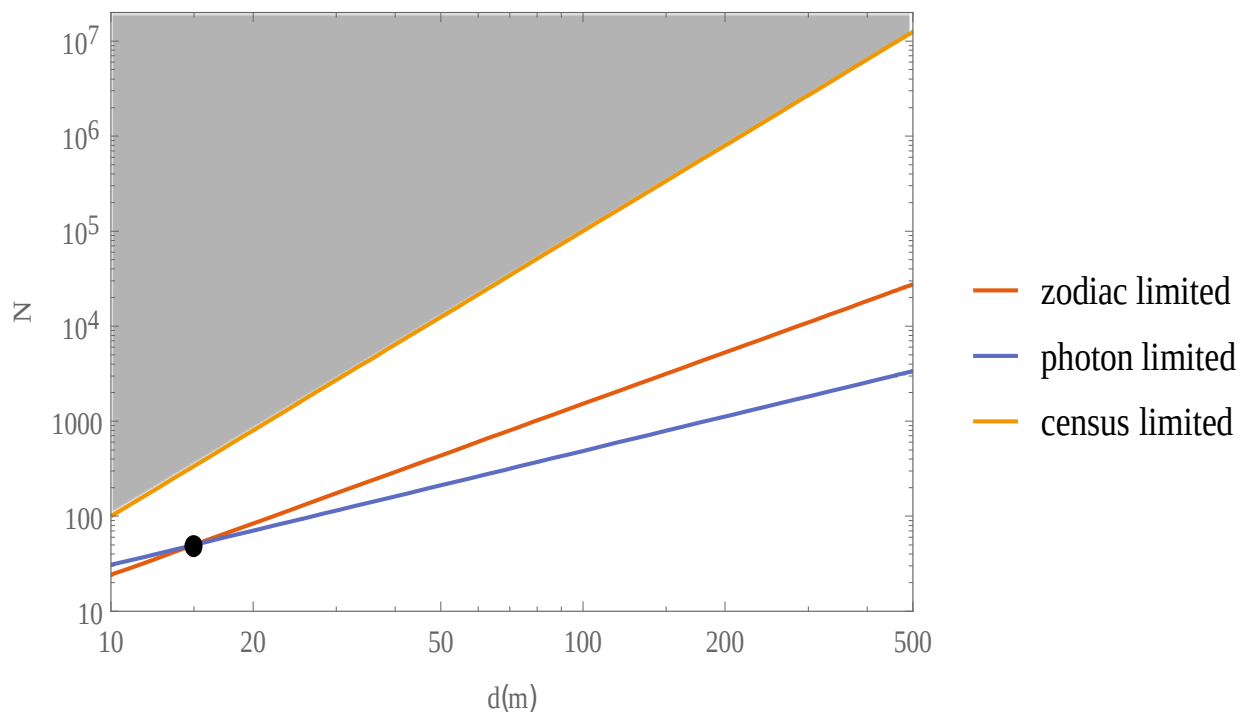

Fig 1: The number of imaged exoplanets as a function of telescope diameter, for various limiting factors, assuming a 20 year mission and extrapolating from the LUVOIR design yield (black dot). The dominant source of noise controlling scaling with telescope diameter is either due to (i) zodiacal dust (red) or (ii) total photon number (blue). The ultimate limit on the total number of stars in volume probed to required resolution is shown in orange as a theoretical upper bound.

The limiting factor, however, may be the contrast ratio: if an orbital starshade is employed, repositioning times may be long. It is also worth mentioning that as opposed to ground-based telescopes situated on Earth, the moon's atmosphere will not introduce a contrast floor, potentially granting access to habitable zones around sunlike stars as well as M dwarfs.

Extension to NIR wavelengths would give access to spectral peaks of many chemical species, including water, methane, carbon monoxide and dioxide, and oxygen. Observed together, these have been argued to indicate the presence of life, and would be feasible with a one hour exposure time at a distance of 10 pc with a 50-100 m telescope (Gaudi et al. 2018), assuming photon-limited noise over one month of observation. Additionally, IR excesses indicating the presence of excess planetary heat are prime candidates for volcanoes and even technosignatures.

Another way of quantifying the yields of different telescope missions is by the certainty with which we can infer the fraction of planets that have various biosignatures. This is typified by the sequence of planets which possess compatibility for life, such as prebiotic chemistry and host star habitability zone, fraction of those which have developed photosynthesis, then multicellularity, and finally technology. The precise biosignatures corresponding to each of these stages of development will not be outlined here, but planets with each of these characteristics are expected to be successively rarer. The extent to which we can infer the fraction that attain each will be highly limited by the sample size of the exoplanet yield. This is displayed in Fig. 2.

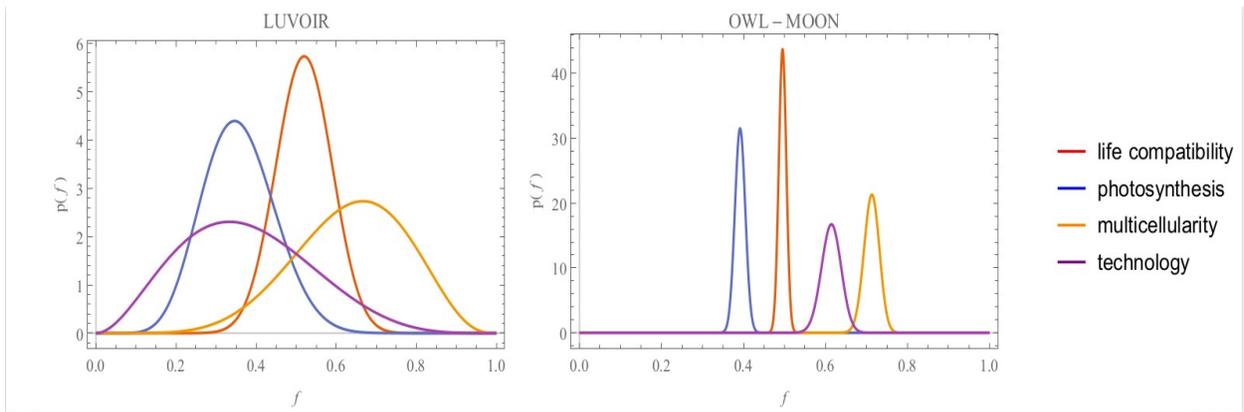

Fig. 2: Inferred values of numbers of planets which attain each successive stage of evolution, assuming optimistically high values of life compatibility fraction $f_L$=0.5, photosynthesis fraction $f_L f_P$ for $f_P$=0.4, multicellular fraction $f_L f_L f M$ for $f_M$=0.7, and technology fraction $f_L f_P f_M f_T$ for $f_T$=0.6. Total sample size for LUVOIR is taken to be 50, and 3,000 for OWL-MOON (McCullen and Silk 2019, in preparation).

## *2b. Stellar physics*

Examples of major contributions to our understanding of stars from ultrahigh resolution spectropolarimetry will include studies of poorly understood atmospheric dynamics including starspots and the role of magnetic fields in the evolution of different types of stars and the launching of stellar winds from massive stars. Spectropolarimetry will enable us to probe stellar environments, ranging from the dynamic upper atmospheres of cool stars to their circumstellar disks and planetary systems. Asteroseismology uniquely explores stellar interiors, and high resolution spectropolarimetry will greatly facilitate this field. Solar seismology is in strong tension with the standard solar model, and it is especially important to develop the field of astroseismology for both main sequence and giant stars to better understand the physics of stellar structure and convective energy transport.

## *2c. Extragalactic objects and cosmology*

One of the goals in the extragalactic domain will be to detect the first stars, galaxies and active galactic nuclei in the Universe at the end of the dark ages, when the very first stars formed. Many telescopes,including the WFIRST and LUVOIR space telescopes, complemented by other terrestrial telescopes, such as E-ELT and SUBARU-PFS, are being planned. However there is no guarantee that their reach will suffice.. We need to look ahead to the future, and develop lunar optical/IR astronomy in order to provide orders of magnitude improvement in sensitivity and in angular resolution relative to all currently envisaged telescopes, whether terrestrial or in space. Our goal will be to motivate use of the ultimate power of telescopes and interferometry on the Moon to help unlock the mysteries of our cosmic beginnings, notably inflation, and to map out the first objects in the Universe. Other goals in cosmology will be to detect the first stars, galaxies and active galactic nuclei at the end off the dark ages.

*2-D images of microlensing*

As an example, let us consider high angular resolution of microlensed quasars. If a point source at a distance DS is perfectly aligned with a foreground source at a distance DL with a mass ML, the latter acts as a « gravitational lens » and creates an annular image (Einstein ring) of the background source with an angular radius $\theta_E = \sqrt{GM_L/c^2 \left(D_L - D_S\right)/D_S D_L}$ and an angular thickness $\theta_S/2$ where $\theta_S$ is the angular diameter of the source in the absence of gravitational lensing. When the foreground lens is off-axis with respect to the lensed source, the Einstein ring breaks into a large and a small arc (see Figure 3).

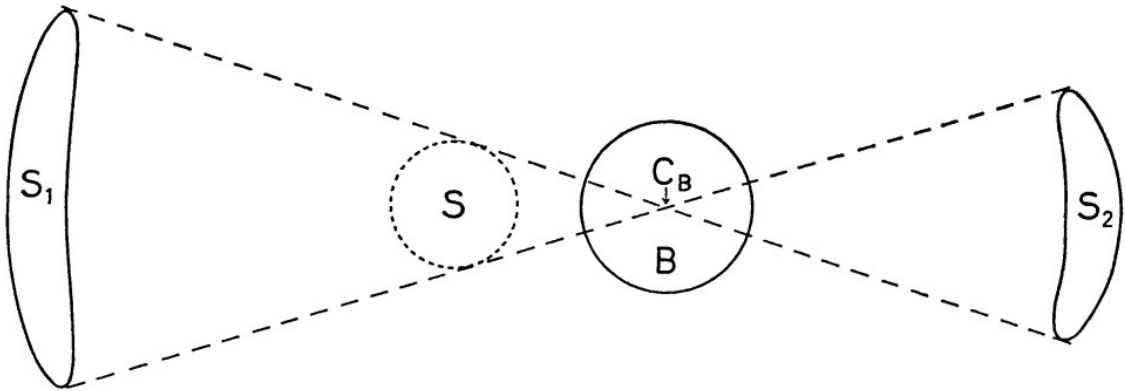

Figure 3: An off-axis foreground lens C breaks the Einstein ring of the background source S into a large and a small arc S1 and S2 (Refsdal 1964).

A stellar type microlensing event has too short a lifetime for its detection by Intensity Interferometry. But extragalactic, most specifically quasars, microlensing lasts for at least decades. In the case of QSO 0957+561 one has DS = 5 Gpc, DL = 1.5 Gpc and ML = $10^9 M_\odot$. From the ratio of 3 of the image brightness BA/BB one infers that the ratio of the length of the two arcs is 3. One could approach the central SMBH via resolving broad line emission regions, at less than 100 mu-arcec angular size, corresponding typically to a scale of 0.2 pc (Sturm et al. 2018). One could map the complex structures of quasar narrow emission line regions at arc-second scales, resolved at ~100 pc scales in nearby AGN.

 Call_for_Voyage_2050_Topical_Team_members.pdf

### 2d. Communication with interstellar probes

One expects that a series of lunar telescopes will have a lifetime well beyond 2050. After 2050, projects for interstellar missions will be well advanced, since the US Breakthrough Starshot Initiative (https://breakthroughinitiatives.org/initiative/3) and in Europe the Initiative for Interstellar Studies (http://i4is.org) are very active on the subject. ESA has also started to identify the problems and their solutions: https://indico.esa.int/event/309/ and the Europlanet program from the European Union is devoting a meeting on the subject: https://meetingorganizer.copernicus.org/EPSC-DPS2019/orals/34097. The problem of how to decelerate the probe once arrived at its destination has already been tackled by European colleagues (Heller et al. 2017). Another problem is how to retrieve information from such a mission. A possible solution is to code the data by bistable meter-wide screens, flipping at several Hertz frequency from opaque to transparent (Schneider & Rigaud, in preparation). When in transit from the parent star of the planet, they will block the stellar light when in the opaque position. With an angular resolution of 60 m. at Proxima Centauri, thanks to the Earth-Moon Intensity interferometer, it creates a detecable dip of ~ 1 % in the signal.

## 3. Additional programmes

As complementary programmes, we propose to use the lunar telescope for the remote sensing of Earth atmosphere and vegetation and as a detector for a « terrascope ».

### 3a. Lunar remote sensing of Earth vegetation and atmosphere

For a while, distant exoplanets (beyond ~ 10-100 pc) will be seen as single dots. To understand their global atmosphere and biosignatures, it will be useful to compare their characteristics with the Earth seen as a whole. Low orbit circumterrestrial satellites, and even geostationary satellites, have too detailed a view to have a global perspective of the Earth. A 1m precursor of a lunar OWL would be useful for global spectro-polarimetric remote sensing of the atmosphere and the surface (land, ocean, vegetation). Also, during lunar eclipses it would allow transmission spectroscopy of the Earth atmosphere blocking the sunlight.

### 3b. « Terrascope » detector

It has recently been proposed to use the Earth atmosphere as a gigantic annular chromatic lens (Kipping 2019). It happens that the focal length of this lens is approximately the Earth-Moon distance, depending on the wavelength. Given the size of this lense, the amplification of the source flux is 20,000 compared to a 1 m telescope (Kipping 2019). With a 100 m telescope on the Moon, the amplification would thus be 200,000 compared to a 30 m telescope. Of course the images are of very poor quality, but this terrascope would be suited for very high spectral resolution or extremely high speed photometry of extremely faint sources (e.g. very faint, yet undetected, optical pulsars). Given the 5° inclination of the lunar orbit with the ecliptic, this terrascope could explore a ± 5° band on the sky above and below the ecliptic, depending on the season.

## 4. Principles and practice of Intensity Interferometry

Intensity Interferometry exploits a second-order effect of light waves by measuring the quantum correlations in light intensity fluctuations from a source that is being observed by two or more separate telescopes. The correlation in light intensity fluctuations, measured as a function of telescope separation (baseline), measures the interferometric squared visibility, which relates to the Fourier transform of the angular brightness distribution of the source across the sky.

The principle underlying intensity interferometry is to monitor the space and/or time correlations of the electromagnetic wave intensity fluctuations in a given, limited bandwidth. The photons are collected either by a single telescope or with two or more telescopes separated by a baseline B at wavelength $\lambda$. Then the achieved angular resolution is $\lambda/B$. If one of the array telescopes is on the Moon, B= 380.000 km on average, corresponding to an angular resolution of 200 pico-arcsecond at $\lambda$=600nm. An Earth-Moon intensity interferometer would partially resolve the Crab pulsar (16.5 visual mag, distance ≈ 2kpc, linear diameter ≈20km), or fully resolve the surface of the brightest white dwarfs such as Sirius B. Attaining such limiting magnitudes (fainter than 15 however ) demands multi-spectral channel intensity interferometry, possibly with 10,000 simultaneous channels of $0.1\overset{\circ}{A}$ resolution. A total spectral bandwidth of 100nm would bring a gain of 5 magnitudes for total exposure times of over ten hours. In practice, it is highly desirable for the Moon telescope to be optically diffraction limited.

Practically, intensity interferometry is insensitive to telescopic optical imperfections, enabling very long (kilometer scale) baselines at short optical (U/V) wavelengths. Furthermore and unlike direct interferometry, intensity interferometry does need optical delay lines to connect the individual telescopes to a common coherent focus. For this to be achieved, photons collected by different telescopes must be tagged and their temporal bunching determined off-line. This needs extremely accurate and stable synchronisation between Earth-Moon telescopes.

Since the whole interferometer will consist, in addition to the lunar telescope, of a few telescopes suitably distributed on the Earth surface, there will be at least one pair of Earth-Moon telescopes operating at any time during the Earth rotation period. It will thus allow continuous monitoring in the intensity interferometry configuration. In addition, thanks to the circulation of the Moon around the Earth, an Earth-Moon interferometer will continuously cover a ring in the (u,v) plane with a +- 5% percent excursion due to the variable distance of the Moon.

Another application of intensity interferometry with quantum optics allows us to distinguish quantum from classical light sources. Detection of time correlations using a single telescope allows investigation of light emission in regions far from thermodynamic equilibrium, such as quasars or other exotic extragalactic objects.

There is already a large amount of experience in intensity interferometry. For instance, it has been succesfully used at the the Narrabri Stellar Intensity Interferometer with two 6.5 m reflectors (Hanbury Brown et al. 1974) and this technique will be used with future ground based Cherenkov Telescope Array (Dravins et al. 2013). In 2019, the VERITAS VHE gamma-ray observatory was augmented with high-speed optical instrumentation and continuous data recording electronics to create a sensitive Stellar Intensity Interferometry (SII) observatory, VERITAS-SII (Kieda et al 2019)a. For reviews, see Rivet et al. (2018) and Kieda et al. (2019b).

Cerenkov intensity interferometry telescopes are best better qualified as photon buckets, The PSF of individual telecopes spread across minutes of arc complicates multispectral channel intensity interferometry which has classical designs with an optical quality telescope. Additionnally, the field of view of such optical telescopes can be split into, for example, 100 sub-fields, and the intensities correlated between corresponding subfields of the two telescopes. This is equivalent to say 100 simultaneuous intensity interferometers that may be applied to detect sub-nano arcsecond structures in the multiple images of lensed QSOs with the Earth-Moon interferometer.

Some science objectives may require more than 100 h exposures. For Intensity Interferometry this is not a problem since one can split them into sub-exposures and accumulate the final total photon counts. Finally, as noted above, Intensity Interferometry is insensitive to the lunar seismology.

## 5. Implementation

OWL-MOON will benefit from the infrastructure afforded by the ESA « Moon Village » initiative. Imagine that we are in the period 2035-2050. By this time many technological advances have been achieved since 2020. For instance, the Nautilus project (Apai et al. 2019) or the WAET project (Monreal et al. 2019) have begun. The Nautilus project has designed new technology for cheap and light 8-meter-class telescopes. This is based on a modified version of Fresnel lenses, made in light plastic. The WAET project is a very large 10 m x 100 m rectangular aperture. The optical quality of these two projects would not be suited for standard interferometry, but suffices for Intensity Interferometry and high resolution spectroscopy.

A complement to interferometry is a monolithic OWL-IR. There are compelling questions that benefit from a very large aperture instrument. The scientific output increases with the effective cosmic volume sampled by the telescope and its detectors, usually as the cube of the effective diameter of the telescope.

We could achieve a 100m optical aperture with an ensemble of off-axis parabolic surfaces. Optically fast parabolic off-axis optics simultaneously provide a narrow field of view for exo-planetary science and a wide field of view for extragalactic astrophysics and cosmology. Maximizing the sub-aperture diameters decreases the overall diffractive light scatter and the complexity cost of the active system. With 8m diameter sub-apertures, the corresponding secondary optics are only a few centimeters across and are small enough to allow active control of the wavefronts from each off-axis sub-unit of the optical system. An optically fast parabolic optics requires a narrow field of view that is less than about 20 microradian.

A 100m diameter will allow statistical searches for life on the nearest 100 or so exoplanets (many of them Earth-like). We will study how to break conventional "moving mass" and complexity cost scaling, create a hybrid "synthetic beam" telescope, and use "no polishing" optical surfaces for a lunar telescope. A proof-of-concept phase will begin with a smaller prototype configuration, based on synergy with the PLANETS telescope project (http://www.planets.life/ ), a 1.9m off-axis telescope under construction at Haleakala, led by a team member, the largest aspect ratio mirror yet used for astrophysical high dynamic range observations.

We will study OWL-IR mirror fabrication from force-sensor-actuator elements that are 3D- printed onto 6-to-10mm-thick slumped glass-sandwich elements of fire-polished glass, and also investigate dust mitigation. The OWL-IR concept is optimized for high angular resolution and high sensitivity in the optical and thermal infrared, as well as high photometric dynamic range, and allows both a narrow-field mode (ELF) optimized for exoplanetary science and a wide-field mode (EELT-like), optimized for extragalactic astrophysics and survey cosmology.

A two-mirror field corrector will be used to optimize corrections across a wide field-of-view, in turn requiring fast optics for an affordable focal plane size, and carry background. The large segmented aperture of OWL-IR and the interferometric arrays share similar requirements in terms of wavefront control, to enable production of the highest quality data by their focal instrumentation. We will study metrology solutions (for co-phasing) as well as enhanced post-processing capability, via the use of generalized closure quantities, so far only exploited in a post-AO context, to more challenging situations such as those of long baseline interferometry on the Moon.

## 6. Precursors

Before building a full OWL-MOON configuration, several 1 to 8 m precursors should be installed on the lunar south pole, to explore early science cases and test intensity interferometry for the brightest sources. They will benefit from previous thoughts in ESA's Directory of Human Flight and Robotics (e.g. Carpenter (2016) and Crawford (2016)).

## 7. Worldwide context

Our OWL-MOON fits well into the « Moon-Village » ESA intitiative which is intended to « bring interested parties together so as to achieve at least some degree of coordination and exploitation of potential synergies » (Woerner 2016). Indeed, the major space agencies of China, India, Japan and USA are already developing activities on and around the Moon. In addition, as an Earth-Moon interferometer, OWL-MOON will have ground-based partners by definition. In particular, after 2035, the European E-ELT will have become more available than during its first years of exploitation and will serve as one of the ground-based counterpart of the Earth-Moon Intensity Interferometer.

The potential use of lunar resources in facilitating the construction of scientific infrastructure on the Moon cannot be underestimated. For example, the HST project was greatly facilitated by the manned space flight program, including development of the space shuttle and construction of the ISS. Similar logic is likely to apply to lunar exploration, with lunar infrastructure likely to subsidize otherwise fiscally unachievable projects such as OWL-MOON.

**Proposing Team**

**Coordinators**

Jean Schneider – Observatoire de Paris (France)

Joseph Silk – IAP (France)/John Hopkins University (USA)

Farrokh Vakili – Observatoire de la Côte d'Azur, Nice (France)

**Co-signers**

Ian Crawford – Birkbeck College of London (UK)

Martin Elvis – Center for Astrophysics, Harvard, Ma, (USA)

Stéphane Jacquemoud - Institut de Physique du Globe de Paris (France)

Robin Kaiser – Université de la Côte d'Azur, Nice (France)

McCullen Sandora - University of Pennsylvania, Philadelphia (USA)